# Transforming the Synthesis of Carbon Nanotubes with Machine Learning Models and Automation

Yue Li,[1, 2, §] Shurui Wang,[1, 3, §] Zhou Lv,[1] Zhaoji Wang,[4] Yunbiao Zhao,[2] Ying Xie,[1] Yang Xu,[2] Liu Qian,[1] * Yaodong Yang,[4] * Ziqiang Zhao,[1, 2] * Jin Zhang[1, 3] *

[1] School of Materials Science and Engineering, Peking University, Beijing, 100871, China.

[2] State Key Laboratory of Nuclear Physics and Technology, School of Physics, Peking University, Beijing 100871, P. R. China.

[3] Beijing Science and Engineering Center for Nanocarbons, Beijing National Laboratory for Molecular Sciences, College of Chemistry and Molecular Engineering, Peking University, Beijing, 100871, China.

[4] Institute for AI, Peking University, Beijing 100871, China.

[§]These authors contributed equally to this work: Yue Li and Shurui Wang.

E-mail: jinzhang@pku.edu.cn; zqzhao@pku.edu.cn; yaodong.yang@pku.edu.cn; qianliu-cnc@pku.edu.cn

## Abstract

Carbon-based nanomaterials (CBNs) are showing significant potential in various fields, such as electronics, energy, and mechanics[1–3]. However, their practical applications face synthesis challenges stemming from the complexities of structural control, large-area uniformity, and high yield. Current research methodologies fall short in addressing the multi-variable, coupled interactions inherent to CBNs production. Machine learning methods excel at navigating such complexities. Their integration with automated synthesis platforms has demonstrated remarkable potential in accelerating chemical synthesis research[4], but remains underexplored in the nanomaterial domain. Here we introduce Carbon Copilot (CARCO), an artificial intelligence (AI)-driven platform that integrates transformer-based language models tailored for carbon materials, robotic chemical vapor deposition (CVD), and data-driven machine learning models, empowering accelerated research of CBNs synthesis. Employing CARCO, we demonstrate innovative catalyst discovery by predicting a superior Titanium-Platinum bimetallic catalyst for high-density horizontally aligned carbon nanotube (HACNT) array synthesis, validated through over 500 experiments. Furthermore, with the assistance of millions of virtual experiments, we achieved an unprecedented 56.25% precision in synthesizing HACNT arrays with predetermined densities in the real world. All were accomplished within just 43 days. This work not only advances the field of HACNT arrays but also exemplifies the integration of AI with human expertise to overcome the limitations of traditional experimental approaches, marking a paradigm shift in nanomaterials research and paving the way for broader applications.

## 0. Introduction

Carbon-based nanomaterials (CBNs), such as carbon nanotubes (CNTs) and graphene, have revolutionized material science with their exceptional electrical, mechanical, and thermal properties[5,6]. From facilitating the fabrication of electronics that surpass the limits of Moore's Law[1], to upgrading the performance of lightweight and high-strength structural-materials[3], to enhancing the efficiency of energy storage[2], CBNs have embarked on a significant journey in advanced materials. However, the full potential of CBNs is often hindered by challenges in synthesizing products with controllable structures, large-area uniformity, and high yield, which are critical for their transition from laboratory research to industrial applications.

This conundrum is a microcosm of the intrinsic challenges prevalent in the development of nanomaterials. Essentially, the journey from atomic assembly to wafer-scale production spans numerous dimensions, exposing the limitations of traditional research methodologies when confronted with such complex systems. Within the conventional scientific paradigm, innovation is often driven by hypothesis-deductive reasoning or analogical reasoning. Hypothesis-deductive reasoning, while powerful in simpler systems, struggles to capture the multi-variable and multi-layered interactions inherent in complex systems like CBNs synthesis. Analogical reasoning, which identifies similar modes and relationships across different systems, is applicable even in high-dimensional complexities and has been widely employed in the development of nanomaterials. However, the success of analogical reasoning heavily depends on extensive expertise, and transferring analogies becomes harder when a field evolves, hindering further breakthroughs.

Once a feasible innovative method for CBNs synthesis is identified, the research focus shifts to optimizing the experimental process and analyzing the outcomes. This phase unveils another challenge that the complex chemical interactions during the synthesis process, involving poorly understood mechanisms, often lead to unexplainable in synthesis recipes and sample performance. Nevertheless, traditional academic optimization and analyzing strategies, which primarily rely on the 'one-factor-at-a-time' (OFAT) method, are inadequate in addressing the coupling effects of various synthesis variables such as catalysts, temperature, and growth substrates, resulting in misunderstandings of the growth mechanisms and overlooking of global optimum conditions[7].

In response to these challenges, there is a pressing need to transform the research paradigm for CBNs synthesis. Machine learning methods, known for their proficiency in navigating the complexities of nonlinear, highly-coupled systems, have emerged as pivotal in this transformation[8]. Together with automation, they herald a new paradigm in scientific research. Recent advancements in materials synthesis, propelled by machine learning and automated experiments, are indeed exhilarating[9]. These advancements spanned diverse areas, including the prediction and synthesis of high-entropy alloys through literature mining[10], the fabrication of nanoparticles by optimization algorithm[11–13], and notably, the recent emergence of an artificial intelligence (AI)-Agent platform for multiple chemical synthesis[14]. The application of this novel scientific paradigm holds tremendous potential in the realm of CBNs.

Therefore, we introduce an AI-driven autonomous chemical vapor deposition (CVD) platform, called Carbon Copilot (CARCO), for the synthesis of CBNs. This platform integrates transformer-based language models, Carbon_GPT and Carbon_BERT, tailored for carbon nanomaterials based on GPTs and BERT, driving innovation in experimental design. Additionally, several data-driven machine learning models are designed to produce specific advice in synthesis process. Complementing these machine learning models is an automated CVD system, which acts as the 'physical extension' of CARCO, enabling around-the-clock experiments and significantly boosting the efficiency and stability of CBNs production.

As a demonstration of the effectiveness of CARCO, our research focused on the synthesis of horizontally aligned carbon nanotube (HACNT) arrays, which are regarded as pivotal materials for advancing next-generation electronics, but encounter bottlenecks in precise synthesis[15]. Over a period of 43 days, CARCO exhibited marked advances in both catalyst innovation and controllable growth of HACNT arrays. Utilizing both Carbon_GPT and Carbon_BERT, we discerned a Titanium-Platinum (Ti-Pt) bimetallic catalyst as a superior and groundbreaking alternative to conventional iron-based catalysts in CNT synthesis. Moreover, we achieved precision in synthesizing HACNT arrays at predetermined densities, making a significant milestone considering the intricate variables involved. These achievements not only advance the CNT development but also robustly validate the role of CARCO in the evolution of CBNs research.

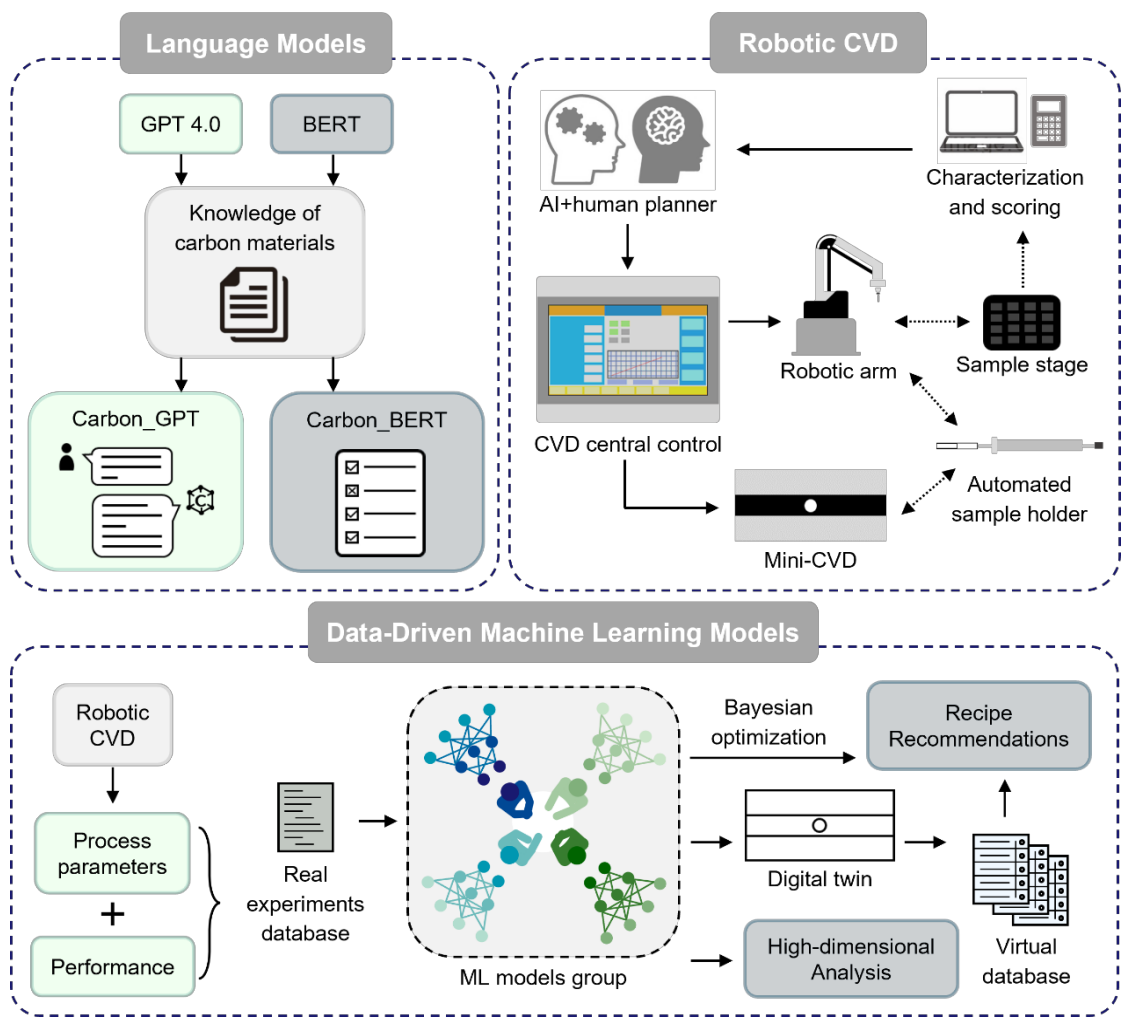

**Fig. 1 | Construction of CARCO platform.** The CARCO platform is composed of three modules: the language models module includes two transformer-based models customized with knowledge of carbon materials for Q&A (Carbon_GPT) and filtering tasks (Carbon_BERT). The robotic CVD module is orchestrated by the CVD central control, solid arrows indicating command flow and dashed arrows showing sample transfer. The data-driven machine learning models module centers around an ML models group built on a database, facilitating process recommendations and high-dimensional analysis tasks.

## 1. CARCO platform

The CARCO platform is structured around three interdependent core components: transformer-based language models[16], robotic CVD, and data-driven machine learning models (Fig. 1). Its modular design enables comprehensive coverage of the CBNs fabrication research process while maintaining flexibility. Human scientists can design workflows targeting specific scientific questions, utilizing various modules as needed.

The transformer-based language models comprise Carbon_GPT and Carbon_BERT. Specifically, Carbon_GPT was created based on OpenAI's custom GPTs[17]. It is constructed by setting specific instructions and uploading a comprehensive knowledge base related to CBNs. This process tailors Carbon_GPT to be adept at identifying and addressing scientific queries associated with carbon materials, thereby offering macro-scale academic insights. On the other hand, Carbon_BERT was produced by fine-tuning of BERT[18,19] with a rich corpus of carbon materials-related literature. Carbon_BERT excels in filtering tasks by utilizing word embedding techniques, such as selecting catalysts for CNT growth.

The Robotic CVD system encompasses several key parts: the CVD central controller, mini-CVD, a robotic arm, an automated sample holder, and the sample stage. The CVD central controller, designed using a programmable logic controller (PLC), executes unified command over the other hardware modules based on experimental parameters ordered by AI or human planners. The system features precise mechanical components for accurate sample placement and retrieval, coupled with meticulous regulation of temperature and gas phases, all situated within a cleanroom environment that maintains constant temperature and humidity levels. This system demonstrates a significant improvement in enhancing the efficiency and consistency of CVD experiments (Extended Data Fig. 1).

For typical synthesis research of CBNs, the robotic CVD system can conduct more than 30 reliable experiments daily. By collecting growth parameters and sample performance analyzed from characterization (Supplementary Fig. 1), it is possible to establish a standardized database comprising over 500 datasets within approximately one month. Leveraging such a high-quality database, we constructed a series of data-driven machine learning models. These models adeptly map the intricate relationships between CVD parameters and the resulting sample performance. In fact, a real-world CVD furnace essentially functions as a predictive model, where a specific set of parameters leads to a distinct performance outcome. Considering the complexity of the parameter space for CBNs synthesis, manually decoding these functional relationships through real-world CVD experiments is highly labor-intensive. Hence, through our data-driven ML approach, we have developed a digital twin of the CVD process, which allows for the simulation of millions of experimental datasets within 20 minutes, identifying appropriate growth parameters by exhaustive search or integrating strategies such as Bayesian optimization[20] (Supplementary Fig. 2). Additionally, these models provide crucial support in unraveling and analyzing the complex interactions between various process parameters and the resultant sample properties.

Emphasizing a collaborative approach, this platform integrates the strengths of AI and human scientists to explore and address intricate scientific challenges in CBNs synthesis. Given that our automated system supports all CVD-based material systems and that the base model tailor approach is highly flexible, coupled with the universality of the data-driven machine learning methods workflow, this platform also holds foreseeable potential in advancing a variety of other nanomaterials.

## 2. Catalyst prediction and high-throughput screening

As the inaugural test of CARCO platform, we embarked on an endeavor to identify innovative methods to the growth of high-density HACNT arrays. Leveraging the advanced capabilities of Carbon_GPT, we explored potential strategies to enhance the density of HACNT arrays. Carbon_GPT's analysis directed us towards several avenues, the optimization of growth conditions, use of tailored catalysts, substrate engineering, and the advance of controlled growth technologies (Extended Data Fig. 2). Among these, use of tailored catalysts, a relatively under-researched area recommended by Carbon_GPT, caught our attention. Historically, iron-based catalysts have been favored for high-density HACNT arrays growth since the early 2000s[21]. The choice of catalysts, while intuitively linked to the structural control of CNTs[22], bears a more complex relationship with array density. Factors like catalyst decomposition capability, carbon solubility, and behavior on substrates contribute to this complexity[23], limiting the scope of exploration in this area for human scientists.

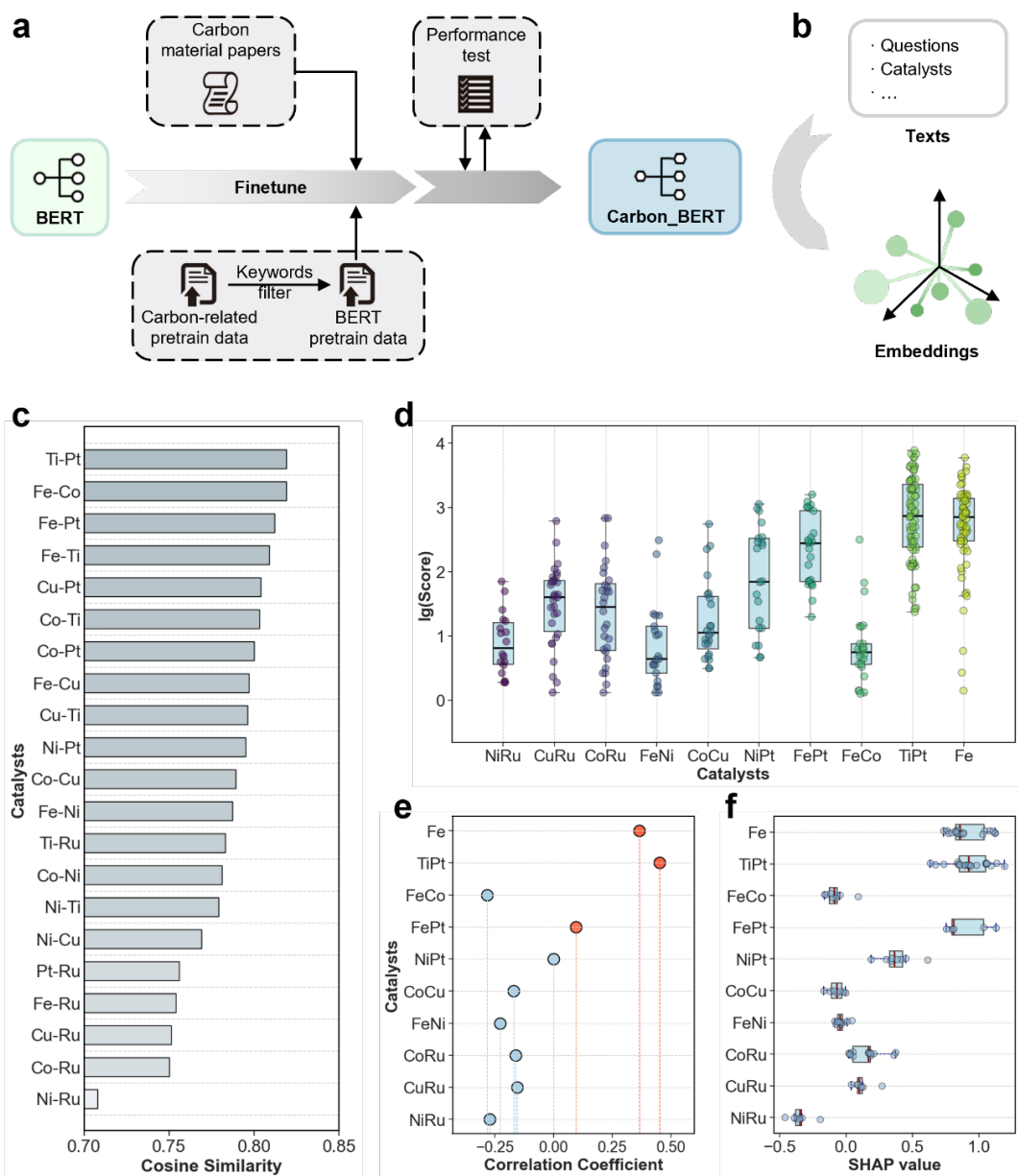

**Fig. 2 | Catalyst prediction and screening.** a, Carbon_BERT's fine-tuning process. b, Schematic representation from texts to embeddings. c, Rankings of cosine similarity for various bimetallic catalysts as derived from Carbon_BERT. d, Density statistics of HACNT arrays with different catalysts. e-f, Spearman correlation coefficients and SHAP values for each catalyst, demonstrating their importance in the synthesis of high-density HACNT arrays, with Ti-Pt showing superior performance over other combinations.

Addressing this multifaceted challenge, we employed a primarily screening approach of catalysts with transformer-based language model, analyzing word embeddings for predictive insights. This method of leveraging word embeddings has already demonstrated significant potential in fields such as thermoelectric material prediction[24] and precursor prediction of inorganic synthesis[25]. Rather than building a model from scratch, we utilized the open-source BERT model, fine-tuning it with carbon material-specific literature, thus creating Carbon_BERT. The training process and results are depicted in Fig. 2a and Extended Data Fig. 3a. Carbon_BERT analyzed the relationships between various catalysts and the concept of 'high-density' in HACNT arrays by transforming texts into word embeddings (Fig. 2b). We ranked potential catalysts based on the cosine similarity between their embeddings and that of 'high-density', predicting the likelihood of their efficacy in enhancing density of HACNT arrays (Fig. 2c, Extended Data Fig. 3b, c and Supplementary Fig. 3). This ranking was not only informed by the specialized knowledge embedded in Carbon_BERT, but also enriched by the diverse insights inherited from BERT's extensive training on a wide range of topics. This method's advantage lies in its ability to tap into the latent knowledge present in the language model, providing innovative suggestions that may not be immediately apparent through conventional scientific exploration.

The predictions regarding single-metal catalysts for CNT synthesis were consistent with the findings of experimental researchers according to previous reports[21,26]. However, due to the scarcity of literature on dual-metal catalysts, a direct evaluation of these predictions was not feasible. Thus, we resorted to high-throughput automated experimentation for both validation and further screening. We selected nine catalyst combinations (three each from the top, middle, and bottom tiers of the prediction list) from the dual-metal predictions, along with conventional Fe as a control. These were loaded onto sapphire substrates using ion implantation technology, followed by the synthesis of HACNT arrays using the automated CVD platform. To ensure a fair comparison among different catalysts, a parameter library was established. This library included recommended parameters based on scientific expertise and randomly generated parameters within a certain range. Each catalyst underwent trials with all parameters in this library, and the order of sample growth was randomized. The samples were then characterized using Scanning Electron Microscopy (SEM), and rated based on the density and orientation of the HACNT arrays. Analysis of the resulting database led to exciting findings, as depicted in Fig. 2d and Supplementary Fig. 4. The performance of most catalysts closely followed the order predicted by Carbon_BERT, except for Fe-Co. Notably, Ti-Pt, ranked first by Carbon_BERT and beyond conventional scientific expectations, outperformed the traditional Fe catalyst in our experimental results.

To further mitigate the influence of catalysts on density performance, the database was analyzed employing both traditional statistical methods and machine learning techniques. Catalysts were encoded using the One-Hot encoder, and Spearman's rank correlation coefficient analysis was utilized (Fig. 2e). P-values for all variables were found to be below 0.05, with the ranking of correlation coefficients displaying high consistency with the average sample performance obtained from experimental observations. Spearman's rank correlation coefficient offers a unidimensional correlation assessment between variables, characterized by strong interpretability. However, it may not fully capture the complexities inherent in the relationships between variables. Consequently, encoded catalysts, in conjunction with the growth process and performance were combined to develop a random forest regression (RFR) model, upon which the SHAP (SHapley Additive exPlanations) method[27] was applied to assess the importance of features for each catalyst (Fig. 2f). The RFR model, adept at processing high-dimensional data and complex feature interactions, in conjunction with SHAP, provides a quantifiable insight into each feature's predictive contribution. The outcomes were largely in agreement with the model's predictive ranking, suggesting that bias towards any specific catalyst was avoided in the experimental parameter settings. Such findings affirm the potential for catalyst prediction innovation through these AI methodologies. The effects of other variables on growth outcomes were also analyzed using these methodologies (Supplementary Fig. 5).

Investigating why TiPt is particularly suitable for the fabrication of high-density HACNT arrays necessitated a collaborative effort between human scientists and CARCO. A thorough literature review revealed that the combination of Pt and $TiO_2$ is a common catalyst in the field of photoelectrocatalysis[28]. $TiO_2$ serves as an excellent carrier for Pt, and together they form an ideal metal-semiconductor interface, exhibiting enhanced catalytic activity in hydrogen evolution reactions. This understanding could be seen as the basis for the model's prediction, suggesting that the innovation stems from the knowledge transfer from the field of photoelectrocatalysis. Experimentally, while human scientists may find it challenging to quantitatively assess samples, they excel at capturing comprehensive information during characterization. For instance, we observed that samples with Pt-based catalysts generally appeared cleaner in SEM characterizations, with fewer bent carbon nanotubes (Supplementary Fig. 6). This observation suggests that the presence of Pt may enhance the utilization efficiency of both catalyst nanoparticles and carbon source. Additionally, TiPt, FePt, and NiPt standing out in the SHAP ranking also indicates the extensive applicability of Pt in these catalyst formulations. However, standalone Pt catalysts failed to produce uniformly distributed HACNT arrays, indicating that the other metal in the combination might play a role akin to $TiO_2$ in photoelectrocatalysis, assisting the behavior of Pt on the substrate and synergistically facilitating the fabrication of high-density HACNT arrays. Interestingly, in terms of interpreting the mechanism, Carbon_GPT offered insights akin to those proposed by human scientist (Supplementary Fig. 7). More detailed studies are currently underway to further unravel these phenomena, with findings to be shared in future publications.

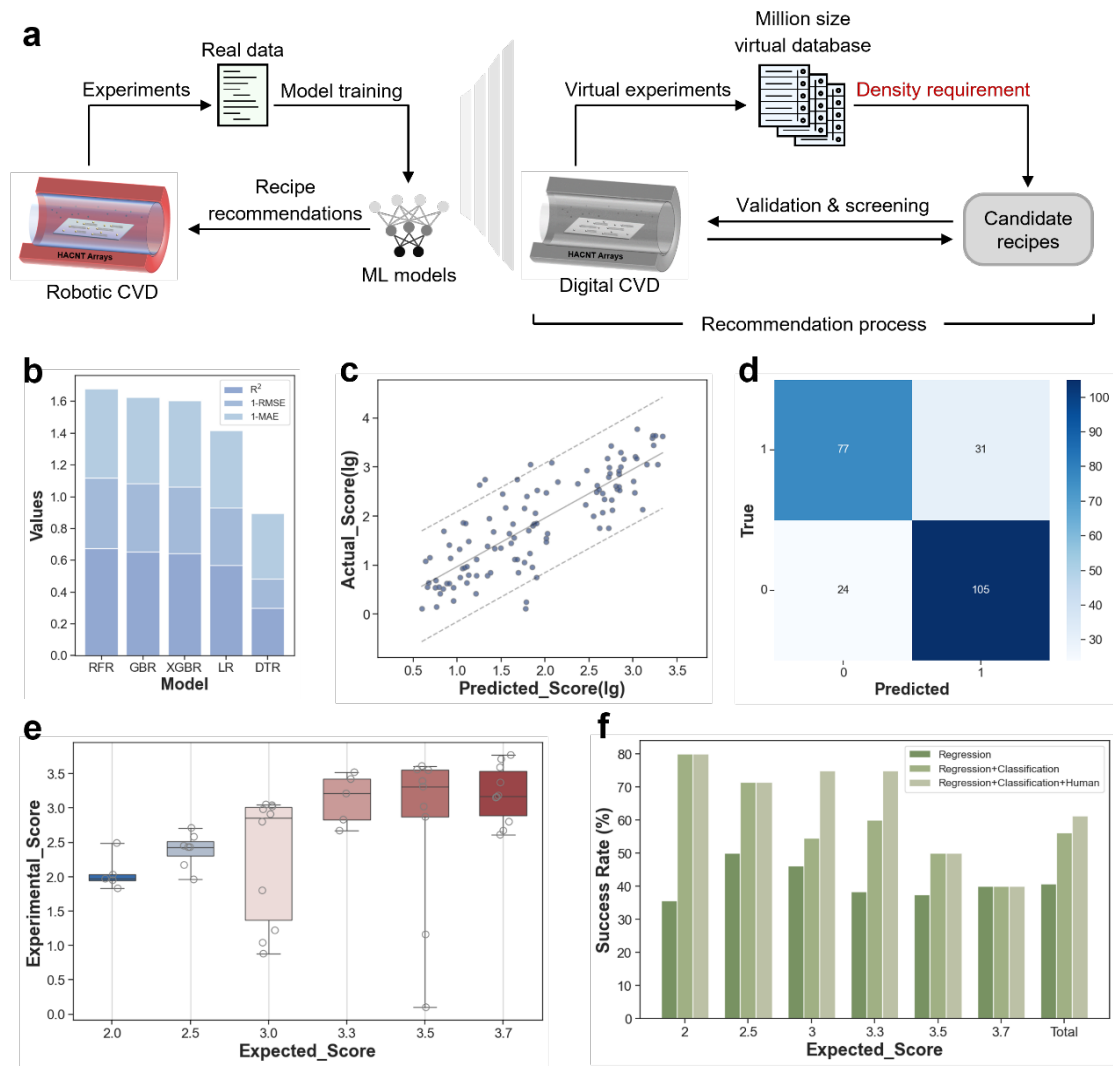

**Fig. 3 | Density-controllable synthesis.** a, Workflow for model establishment and density-controllable synthesis. b, Performance metrics of regression machine learning models. c, The RF model's performance on the test set, with a scatter plot showing the correlation between actual and predicted HACNT density scores. d, Confusion matrix for the classification accuracy of growth occurrence. e, Experimental validation of density-controllable growth, with the x-axis representing specified densities and the y-axis showing results from real-world experiments. f, Comparative success rates of different approaches: solely regression, regression combined with classification, and incorporating human evaluation.

## 3. Density-Controllable Synthesis

Having demonstrated the system's innovation in catalyst prediction and high-throughput screening, another intriguing question for us was its capability to achieve precise synthesis. HACNT arrays hold extensive potential in fundamental property research and applications in electronic and optoelectronic devices[29], and biomedical sensors[30]. In carbon-based FETs, HACNT arrays with higher density are often sought[31], while in biomedical sensors and fundamental spectroscopy research, maintaining density within the resolution limits of analytical methods is crucial[32]. Therefore, the ability to customize samples with specific densities is an overwhelming superiority of CVD synthesis and has significant application value.

The key to this challenge lies in establishing the functional relationship between process parameters and sample density. Machine learning methods were employed to handle these multifactorial, nonlinear, and non-monotonic complex relationships. We developed a standardized database using CARCO, comprising over 500 data points including catalysts, growth parameters, and characterization results. Multiple machine learning models were built to represent the functional relationship between process parameters and predicted sample scores. In essence, these models acted as digital twins of real-world CVD equipment (Fig. 3a). In the digital twin, we were able to perform up to one million experiments in 20 minutes, facilitating a more nuanced exploration of the CVD growth parameter space.

The quality of these models directly influences the accuracy of the digital twin experiments. Given that the density of HACNT arrays in the parameter space exhibits a log-normal distribution, we first eliminated samples with a score of 0, then applied a logarithmic transformation to the remaining scores. Bayesian optimization was utilized to find the best parameters for the models, with the Random Forest (RF), Gradient Boosting Regressor (GBR), and XGBoost (XGB) models emerging as winners, achieving $R^2$ scores of 0.67, 0.65, and 0.64 respectively (Fig. 3b, c and Supplementary Fig. 8). Given the inherent randomness in CVD experiments and score assessments, the interpretability of the models for real experimental data was quite satisfying. Predicting the density of HACNT arrays involves both classification (whether growth occurs) and regression (how dense the growth is) problems. Our initial data engineering led to a lack of understanding of the non-growth parameters in regression, thus we established a separate classification model as a filter to exclude non-growing parameters suggested by the regression models (Fig. 3d). Consequently, a digital twin comprising four models was constructed, generating a virtual database of experiments on the scale of millions.

Based on this, we developed a workflow for fabricating CNT arrays with specified densities, as depicted in Fig. 3a. To achieve a certain density, we simply need to find the closest-matching process parameters in the virtual database. Using an ensemble training-like method, we cross-validated the parameters suggested by various models, assigning a credibility score to each set. Finally, the classification model filters out parameters unlikely to result in growth, yielding candidate sets (Supplementary Table 1). Fig. 3e and Extended Data Fig. 4 show the experimental validation of this process.

A systematic experimental validation was conducted across a spectrum of densities, ranging from 0.5 to 25 nanotubes per micrometer (lg(score) ranging from 2 to 3.7), corresponding to scores from 2.0 to 3.7. With a tolerance margin of 10% in lg(score), the process parameters recommended through the established workflow demonstrated a striking accuracy rate of 56.25% (27/49). This efficacy distinctly surpasses the 39.74% (31/79) accuracy achieved when solely dependent on regression models.

Human brains excel in performing tasks akin to a classification model[33]. We had human scientists filter parameters suggested by regression models, achieving an accuracy of 49.15% (29/60), which is lower than that of classification models. Interestingly, when human scientists judged parameters refined through a secondary screening by classification models, the accuracy reached 61.36% (27/45). Human scientists successfully eliminated four sets of parameters that did not meet the specified density targets, reaffirming the synergy between human intuition and CARCO suggestions. Figure 3f presents the success rate of parameter recommendations for each density. At lower densities, the candidates provided by regression models contained many that were unviable for growth, likely due to more distinct features, allowing both classification models and human scientists to effectively eliminate poor parameters. At higher specified densities, the parameters suggested by regression models were generally reasonable, making it challenging for classification models and human scientists to identify non-viable parameters. Theoretically, the results mentioned above could be further improved with the expansion of the database. In summary, with the assistance of the CARCO platform, we can achieve precise control that is unattainable with traditional methods, opening up possibilities for the application of HACNT arrays in a broader range of fields.

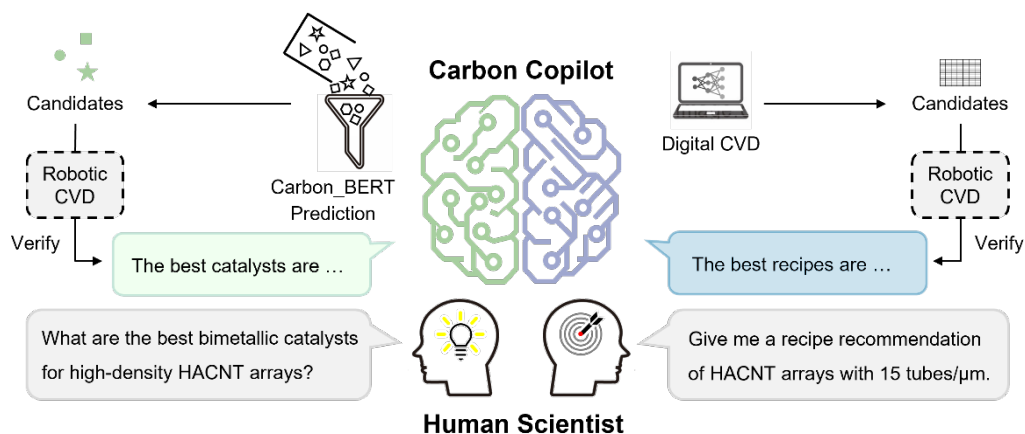

**Fig. 4 | Interactive workflows with CARCO.** Two workflows, catalyst prediction (left side) and controlled-density growth (right side), demonstrating CARCO's capabilities in innovation and precision manufacturing within CBNs synthesis research.

## 4. Conclusions and Outlook

Our study made valuable contributions to the field of HACNT arrays. By leveraging various modules of CARCO, we have not only achieved catalyst prediction for HACNT arrays synthesis, corresponding to an innovative challenge, but also accomplished the targeted density growth of HACNT arrays, corresponding to an optimization challenge. Fig 4 illustrates the workflows for accomplishing these two challenges. A noteworthy aspect of our work is the unexpected alignment between Carbon_BERT's catalyst predictions and the actual experimental results, highlighting the efficacy of embedding methods in the synthesis of CBNs. This alignment led to the innovation of Ti-Pt as a potential dual-metal catalyst, which exhibits promising attributes compared to traditional iron catalysts for the synthesis of high-density HACNT arrays. Moreover, our platform has enabled the synthesis of HACNT arrays with specified densities, achieving a precision rate of 56.25%, which is a "mission impossible" without AI methods. It's worth noting that all the experiments were completed in approximately one and a half months. The current efficiency is primarily constrained by the efficiency of nanomaterial characterization.

The vital role of machine learning in advancing the CBNs field has been underscored by this research. A direct advantage observed is the marked increase in experimental throughput and consistency achieved with the Robotic CVD, particularly crucial for addressing the complex system challenges inherent in nanomaterial synthesis. The potential of transformer-based language models to inject innovation into nanomaterial research has been validated, possibly due to their ability to facilitate cross-disciplinary creativity, supported by their extensive and diverse training data. Diverging from other recent "AI for Science" publications[9,14,20], this work emphasizes the collaborative synergy between AI and human scientists. This exploration of cooperation is particularly meaningful for nanomaterial research, where characterization and evaluation pose complex challenges. In fact, compared to traditional macroscopic materials, the characterization of nanomaterials presents a higher level of complexity. For example, in our case, for the density of HACNT arrays, it is difficult to obtain high-quality quantitative results with high-throughput and automation-compatible spectroscopic characterization methods, requiring the assistance of human scientists for SEM characterization and measurement. We believe it is important to explore the cooperative pathways between AI agents and human scientists. For the time being, at least, CARCO acts as a copilot for human scientists, not a replacement.

Looking towards the future, our study's approach holds immense potential for scalability and adaptability across various nanomaterial systems. The flexibility of our method, anchored in tailoring of transformer-based language models to specialized domain-specific models, offers a robust framework that can be applied to other domains. The automated CVD platform is naturally suitable for a wide range of nanomaterials like graphene and molybdenum disulfide. Building upon these foundations, the system aims to reshape perspectives in nanomaterial synthesis. The advent of AI technology is set to revolutionize research methodologies. The efficacy of AI in this context is not solely in its standalone capabilities but significantly in its synergistic collaboration with human scientists. We hope our research can serve as a prototype, demonstrating the profound impact of AI technologies in advancing the nanomaterials field.

# Methods

**Construction Details of CARCO**

To construct the Carbon_GPT for our CARCO platform, we tailored a GPT specifically for the domain of carbon-based nanomaterials (CBNs), with a particular focus on carbon nanotubes (CNTs). The creation of Carbon_GPT began with configuring specific instructions and incorporating an extensive knowledge base, selected meticulously to encompass the broad spectrum of research and applications pertaining to CBNs. The primary instructions are as follows:

"Carbon GPT, while an expert in carbon materials, focuses predominantly on carbon nanotubes. It offers extensive knowledge from both English and Chinese literature, adapting its responses according to the query's language. The AI's primary functions include: 1) Providing overviews and macro-level guidance on carbon nanotube research, summarizing scientific inquiries, and recommending exploration areas. 2) Detailing preparation methods, applications, and practical advice specifically for carbon nanotubes. 3) Analyzing real-world experimental results related to carbon nanotubes, interpreting data, and offering insights for improvement. While its expertise extends to broader carbon materials, Carbon GPT's core strength and focus remain on carbon nanotubes, ensuring in-depth, tailored responses in this area."

Carbon_BERT underwent fine-tuning with a comprehensive corpus of literature on carbon materials. On one hand, we selected high-quality literature related to carbon materials and utilized GPT-4 for data cleaning; on the other hand, we used a search engine to filter out texts related to carbon materials and catalysts from the original BERT training data. Together, these formed a knowledge base on carbon materials. In processing the data for fine-tuning, we paid special attention to the structure of the input text. We ensured that distinct lines of text in the dataset were handled as distinct sequences, which is crucial for models designed to understand and generate text based on complex scientific literature. This strategy allowed for a more nuanced and contextually aware learning process, particularly beneficial for the specialized domain of carbon materials.

During the fine-tuning process, we employed the masked language model (MLM) loss function, foundational to BERT's training. By randomly masking tokens in the text and predicting these based on the context of surrounding tokens, we deepened the model's grasp of language structures and domain-specific terminologies. Adjustments to the proportion of masked tokens and the exploration of different learning rates further optimized the model's ability to internalize the intricate vocabulary and concepts central to carbon materials.

To assess the effectiveness of our training approach, we devised 12 specific test questions covering a broad spectrum of topics within the carbon materials field. The outcome was a notable improvement in the model's performance, evidenced by the increase of the Spearman correlation coefficient from 0.1 to 0.3.

The Robotic CVD system encompasses a CVD central controller, a mini-CVD, a robotic arm, an automated sample holder, and a sample stage. The mini-CVD selected is the Micro-STS1200 from Units Technology Co., Ltd. Its compact design and pre-installed optical windows ensure the scalability of the Robotic CVD system. Custom sample rods were designed and connected to linear rails, with the sample stage being moved in and out by stepper motors. Optical sensors were installed on both sides for positioning, and a specially designed wedge-shaped furnace opening ensures a good seal under certain pressure conditions. The robotic arm selected is the MG400 from Dobot, is used for placing and picking up substrates with a suction nozzle. All modules are centrally controlled by the CVD central controller, which is equipped with a PLC (Panasonic C40ET). It can be operated through a home-made interface or connected to a personal computer (PC) for control.

In constructing the data-driven machine learning models, we initiated our framework with data engineering. This involved the preprocessing of the dataset where catalyst variables were transformed to one-hot vectors to convert categorical data into a machine-readable format. The synthesis parameters were retained in their original form to maintain the integrity of the experimental conditions. The response variable, 'Score', representing the performance of HACNT arrays, was logarithmically transformed to lg(Score) to normalize the distribution of the data and stabilize the variance. The dataset was split into training (0.7) and testing (0.3) subsets to ensure the robustness and generalizability of the model. Subsequently, RFR, GBR, XGBR, LR and DTR were developed, respectively. We employed Bayesian optimization as a strategic approach to refine the hyperparameters of the model. This process began by defining a range for each hyperparameter, which guided the optimization algorithm. The Bayesian optimizer was then initialized to explore the hyperparameter space, balancing the trade-off between exploration of new parameters and exploitation of known to be effective ones. The objective function to maximize was the cross-validated performance of the model on the training set. The output of the Bayesian optimization provided a set of best hyperparameters, which were used to train the final ML models. The performance of the models was evaluated using the $R^2$ score, Mean Absolute Error (MAE), and Root Mean Square Error (RMSE) on the testing set. These metrics provided insights into the accuracy and reliability of the model's predictions (Supplementary Fig. 9).

In addition to regression analysis, we also implemented a binary classification to distinguish between growth and non-growth samples. This was achieved by creating binary labels from the 'Score' with a defined threshold. A pipeline was established to streamline the process of predicting the likelihood of growth occurrence. The process concluded with an evaluation of the model's performance, ensuring the accuracy and efficacy of the predictions (Supplementary Fig. 10).

**Catalyst Screening**

In the initial phase of catalyst screening, we established a comprehensive list that consisted of a series of questions along with a list of potential candidate catalysts. For the processing of our textual data, the BertTokenizer from the transformers library was employed to convert the text into a format comprehensible by the model. The BertModel, a PyTorch implementation of BERT, was utilized to operationalize the pre-trained Carbon_BERT model. This setup facilitated the conversion of all listed words into embeddings, effectively capturing the contextual relationships inherent in the language model's training data. The embeddings for each word in the list were generated by passing the text through the tokenizer and subsequently through the model, focusing on extracting the last hidden state of the [CLS] token representation, which serves as an aggregate embedding for the input sequence.

Upon obtaining the embeddings, the cosine similarity between the question embedding and each of the candidate catalyst embeddings was computed. This metric evaluates the cosine of the angle between the vectors in the embedding space, generating a similarity matrix. The similarity scores specific to the predefined query regarding high-density SWNTs synthesis were extracted, and the candidate catalysts were ranked according to their cosine similarity to the query. This ranking was indicative of the likelihood of each catalyst's efficacy in enhancing the density of SWNTs, thus guiding the selection process towards the most promising candidates for further experimental validation.

**Growth and Characterization of HACNT Arrays**

We utilized a-plane sapphire substrates (single-side polished, miscut angle < 0.5°, surface roughness < 0.5 nm) for the preparation of HACNT arrays, which were acquired from Hefei Kejing Materials Technology Co., China. The initial stage involved loading the catalysts onto the substrates using ion implantation technology conducted at room temperature. This implanter (FAD-MEVVA) is equipped with a uniform beam spot exceeding 8 inches in diameter[34], a specification verified using Gafchromic EBT3 self-developing dosimetry film[35]. Such uniformity enables the concurrent implantation of catalysts into five 3-inch sapphire wafers, with each wafer segmented into roughly 160 samples measuring 4cm x 6cm. As a result, each implantation cycle can consistently deposit catalysts on approximately 800 samples. The ion fluence was controlled within a range from 1E13 to 1E16 ions/cm², and the ion energies were adjusted between 5 to 20 keV. We employed a range of elements including Fe, Co, Ni, Cu, Ti, Pt, and Ru. The physical phenomena of ion implantation were examined using The Stopping and Range of Ions in Matter (SRIM) software suite. Specifically, the Monte Carlo-based Transport of Ions in Matter (TRIM) simulation was implemented to probe the interactions between energetic ions and the sapphire substrates[36].

After the ion implantation, the samples were sequentially ultrasonically cleaned in deionized water, acetone, and ethanol. An annealing process at 1100°C in an air atmosphere for 5 hours was then conducted. This step was crucial in mending the radiation-induced damages to the substrate surface from the implantation.

The growth process was fully automated by the Robotic CVD system. Substrates were conveyed into the mini-CVD by the robotic arm and automated sample holder. Upon the initiation of the growth cycle, the system was programmed to reach the predetermined temperature in 15 minutes. Once the target temperature was achieved, the system sequentially executed the reduction and growth phases. Following the growth cycle, the temperature was rapidly reduced by water-cooling, taking approximately 7 minutes for the system to reach 150°C. At this point, the samples were automatically retrieved by the sample holder and robotic arm, and the system proceeded to the next set of growth tasks.

The parameter space for the synthesis was defined with specific ranges to optimize the conditions for HACNT array growth. The parameters were set as follows: Temperature: ranging from 800°C to 1000°C; Reduction time: ranging from 1 to 1000 seconds; Growth time: ranging from 1 to 1000 seconds; Argon flow: ranging from 50 to 500 sccm; Hydrogen flow: ranging from 15 to 500 sccm; Ethanol flow: bubbling by argon flow, ranging from 1.7 to 118.3 sccm.

**Performance Evaluation of HACNT Arrays**

We performed general characterization of HACNT arrays, including SEM, Raman, and AFM analyses (Supplementary Fig. 11). SEM images were obtained on a Hitachi S4800 SEM operated at 1.0 kV and 10 kV. Raman spectra of SWNTs with line mapping conducted with a step of 5 μm and a laser beam spot of ~1 μm were collected from Jovin Yvon-Horiba LabRam systems with 532 nm excitation. AFM images were obtained using a Dimension Icon microscope (Bruker).

The score of HACNT arrays is composed of two aspects: density and orientation. Density is computed using a Python-based recognition program that was custom-developed for this purpose, while the orientation is subjectively evaluated to obtain a penalty factor. Scores for each sample are then determined through a standardized computational method.

## Data/model/code availability

The data/model/code will be openly released after peer review.

## Acknowledgements

This work was financially supported by the Ministry of Science and Technology of China (2022YFA1203302, 2022YFA1203304 and 2018YFA0703502), the National Natural Science Foundation of China (Grant Nos. 52021006, 52102032, 52272033), the Strategic Priority Research Program of CAS (XDB36030100), the Beijing National Laboratory for Molecular Sciences (BNLMS-CXTD-202001) and the Shenzhen Science and Technology Innovation Commission (KQTD20221101115627004).

## Author contributions

Y.L. and S.W. prepared the samples and performed the experiments of synthesizing and characterizing HACNT arrays, assisted by Y.Z., Y.X., and Y.X. Y.L. performed the design and construction of the CARCO platform, Z.W. conducted the fine-tuning of language models, supervised by Y.Y. Y.L., L.Q., and S.W. wrote and revised the manuscript with input from all authors. J.Z., L.Q., Z.L., and Z.Z. supervised the overall projects. All authors contributed to the discussion and completion of this manuscript.

## Competing interests

The authors declare no competing interests.

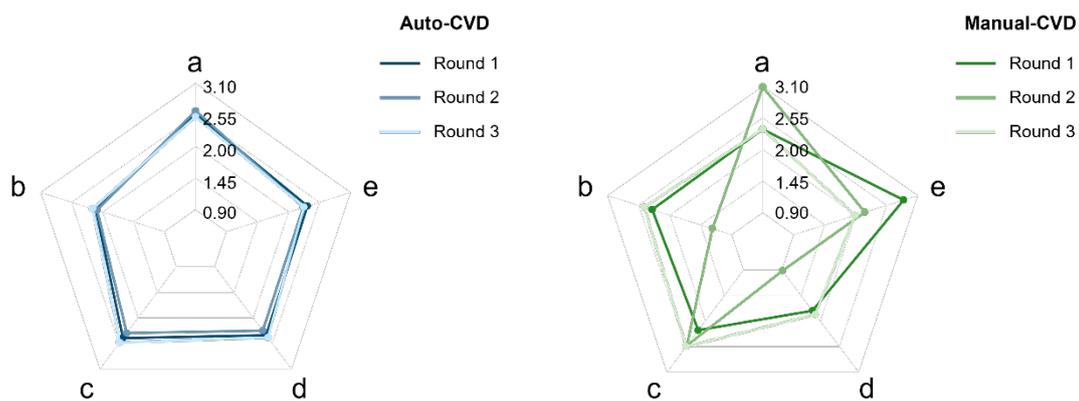

|   | cat. type | T/°C | $H_2$/sccm | EtOH/sccm | Ar/sccm | red_t/s | grow_t/s |
|---|-----------|------|------------|-----------|---------|---------|----------|
| a | Fe 6E15   | 930  | 200        | 6.9       | 111     | 360     | 780      |
| b | Fe 6E15   | 930  | 300        | 6.9       | 111     | 120     | 780      |
| c | Fe-Pt     | 930  | 100        | 6.9       | 111     | 240     | 780      |
| d | Fe 1E16   | 910  | 100        | 3.3       | 300     | 120     | 780      |
| e | Fe 6E15   | 930  | 100        | 3.3       | 300     | 120     | 780      |

**Extended Data Fig.1** | Experimental reproducibility assessment. Five different sets of catalysts or growth parameters, labeled a-e, underwent three rounds of experimentation using both robotic CVD and traditional manually-controlled CVD setups. The robotic CVD demonstrated a significant consistency advantage, showcasing the reliability and repeatability of the automated system.

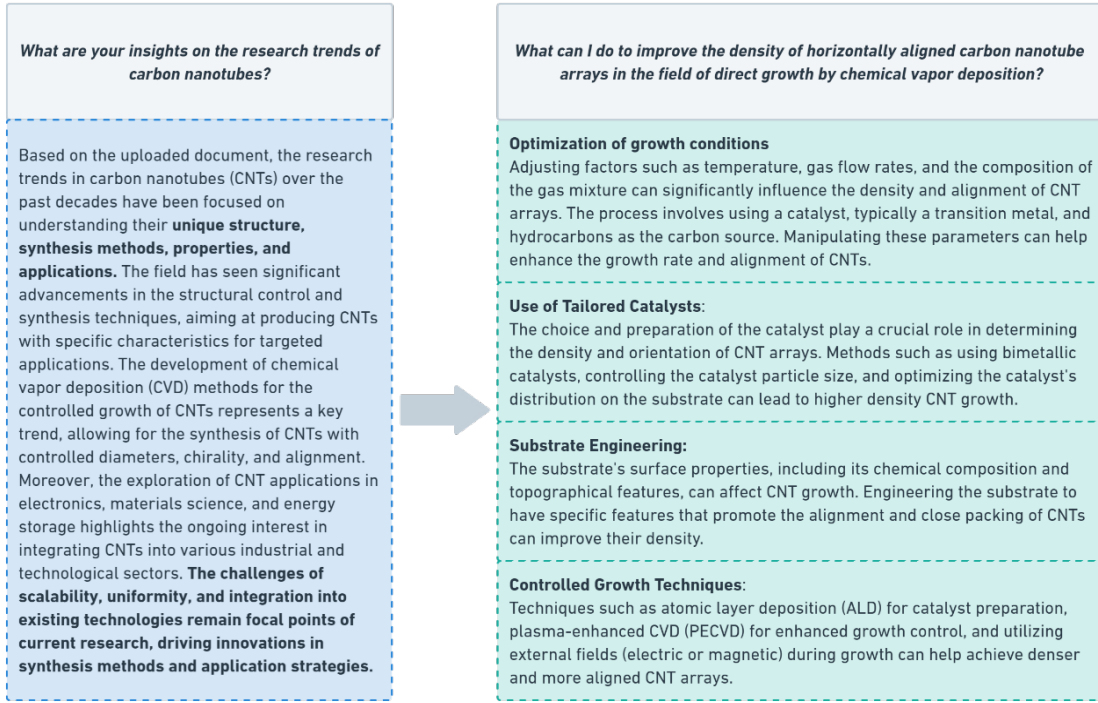

**Extended Data Fig.2** | General suggestions by Carbon_GPT. The blue area represents insights into the developmental trends of carbon nanotubes. The green area provides suggestions on enhancing the density of HACNT arrays, including the use of tailored catalysts, with a specific mention of bimetallic catalysts.

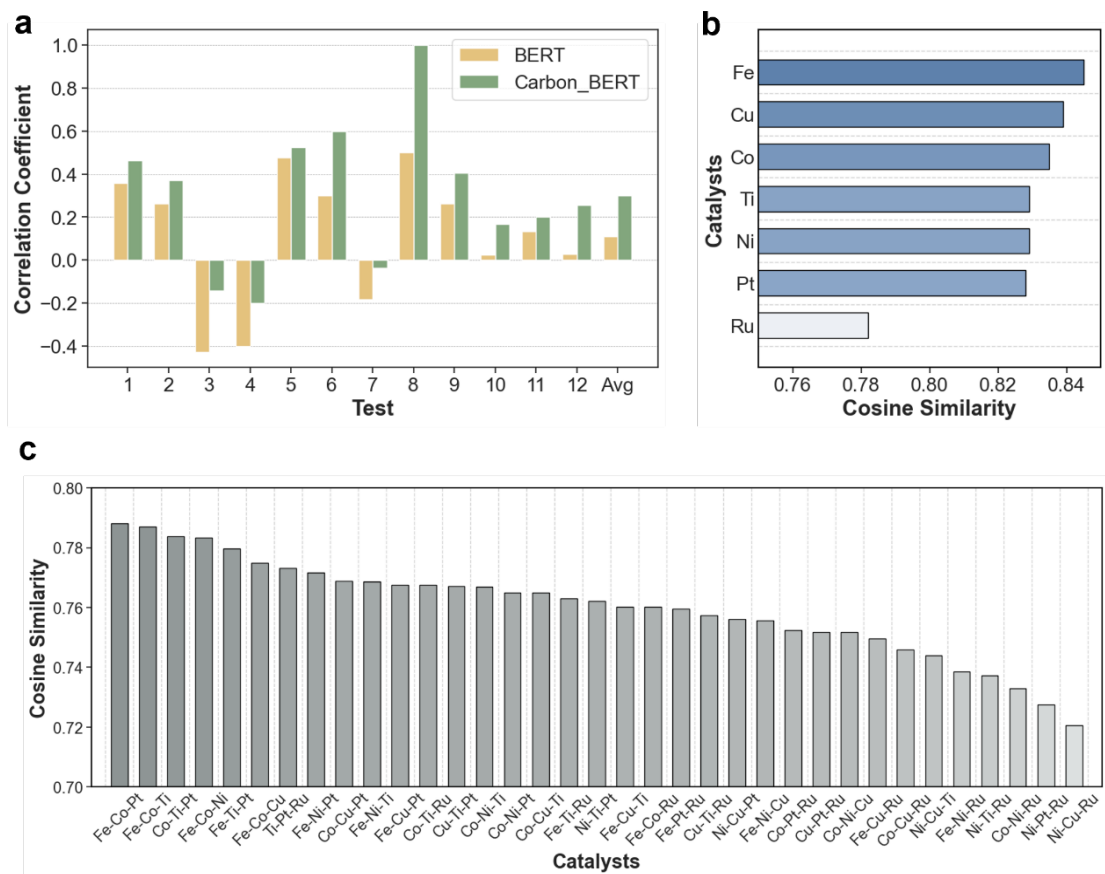

**Extended Data Fig.3** | Evaluation of Carbon_BERT. a, Performance improvement of Carbon_BERT over BERT on a set of test questions. b-c, Recommendations by Carbon_BERT for the preparation of high-density HACNT arrays on single and ternary metal catalysts, respectively.

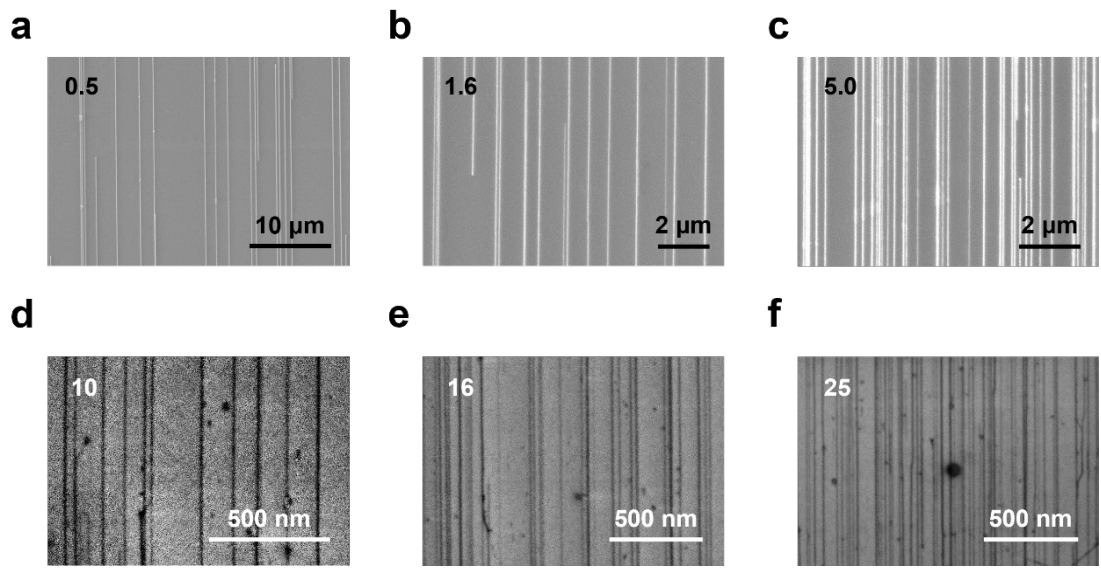

**Extended Data Fig.4** | SEM Images of density-controlled growth. The top left corner of each image marks the specified density targets for the HACNT arrays.

# Supplementary Information

# Transforming the Synthesis of Carbon Nanotubes with Machine Learning Models and Automation

Yue Li,[1,2,§] Shurui Wang,[1,3,§] Zhou Lv,[1] Zhaoji Wang,[4] Yunbiao Zhao,[2] Ying Xie,[1] Yang Xu,[2] Liu Qian,[1]* Yaodong Yang,[4]* Ziqiang Zhao,[1,2]* Jin Zhang[1,3]*

[1] School of Materials Science and Engineering, Peking University, Beijing, 100871, China.

[2] State Key Laboratory of Nuclear Physics and Technology, School of Physics, Peking University, Beijing 100871, P. R. China.

[3] Beijing Science and Engineering Center for Nanocarbons, Beijing National Laboratory for Molecular Sciences, College of Chemistry and Molecular Engineering, Peking University, Beijing, 100871, China.

[4] Institute for AI, Peking University, Beijing 100871, China.

[§]These authors contributed equally to this work: Yue Li and Shurui Wang.

E-mail: jinzhang@pku.edu.cn; zqzhao@pku.edu.cn; yaodong.yang@pku.edu.cn; qianliu-cnc@pku.edu.cn

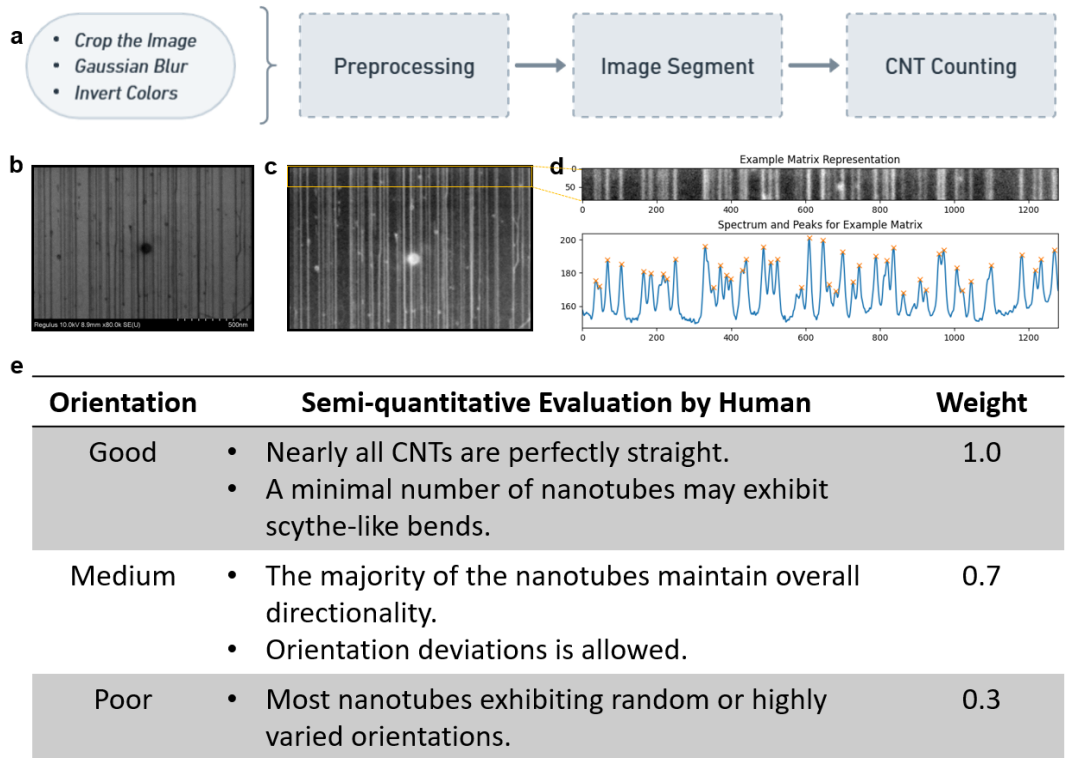

$$Score = Density * Orientation * Constant$$

**Supplementary Fig. 1 | Performance evaluation details of HACNT arrays.** a, Flowchart of the automated program for calculating density from SEM images. b-d, A typical example of density calculation. e, Criteria for semi-quantitative evaluation standards based on CNT orientation, detailing good, medium, and poor classifications with associated weights, and the formula for calculating the overall score.

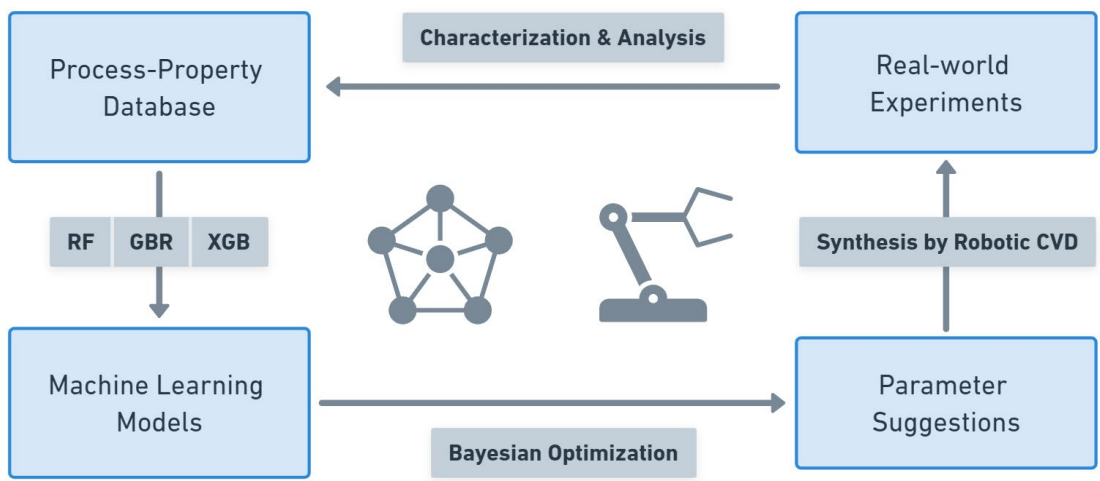

**Supplementary Fig. 2 | Active Learning Workflow of CARCO.** It begins with the construction of machine learning models from an initial database. The models guide Bayesian optimization to propose new experimental parameters. These parameters are then employed in automated experiments conducted by the Robotic CVD system. Following experimentation, the sample is characterized and the new data is integrated back into the database, completing the feedback loop and refining the process.

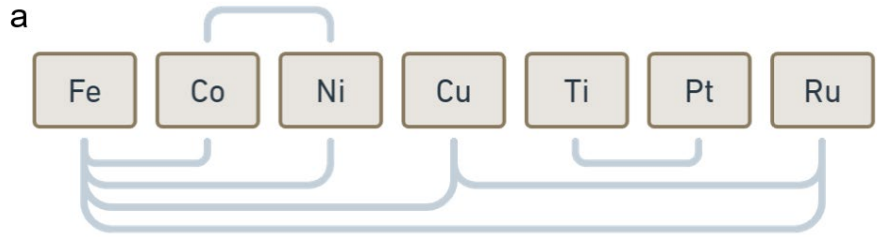

**Supplementary Fig. 3 | Catalyst Combination and Screening.** a, Illustration of the 127 possible combinations arising from seven ion-implantable elements, indicating a need for high-throughput screening methods. b, Displays the top three combinations for four and five metals as ranked by Carbon_BERT, which will be subject to experimental validation in future work.

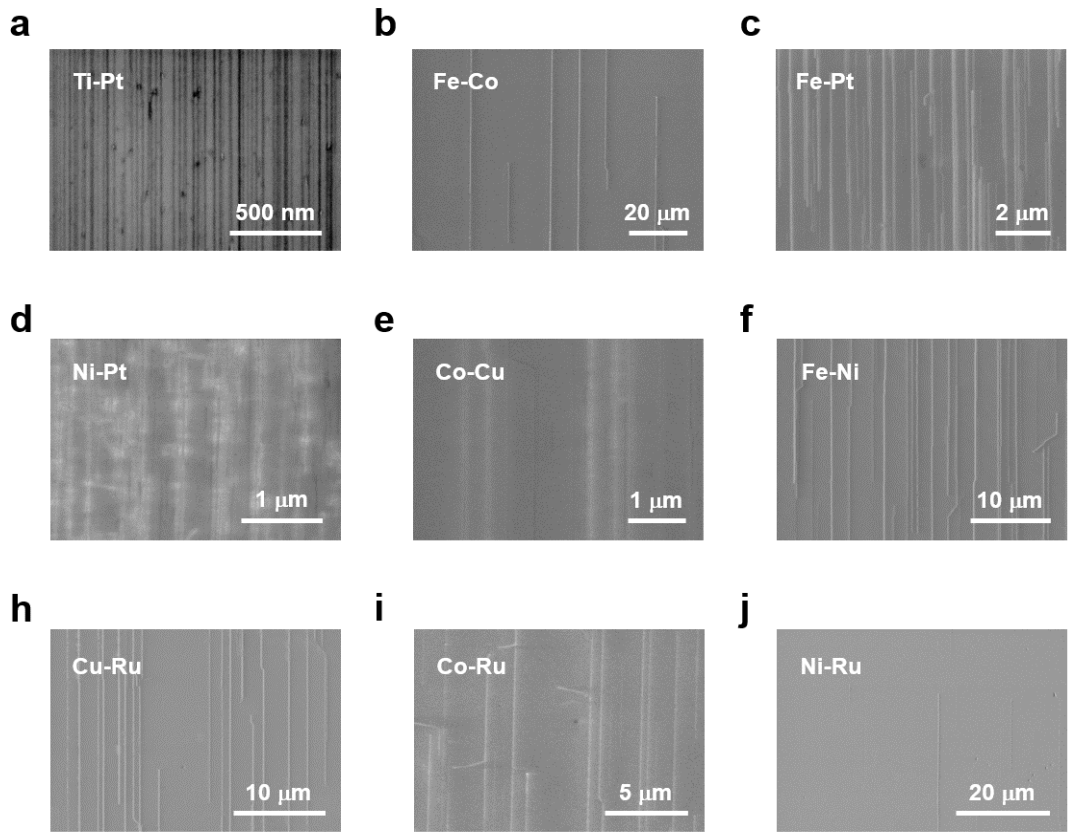

**Supplementary Fig. 4 | SEM images of HACNT arrays growth with various catalysts.** The top left corner of each image marks the specified catalyst.

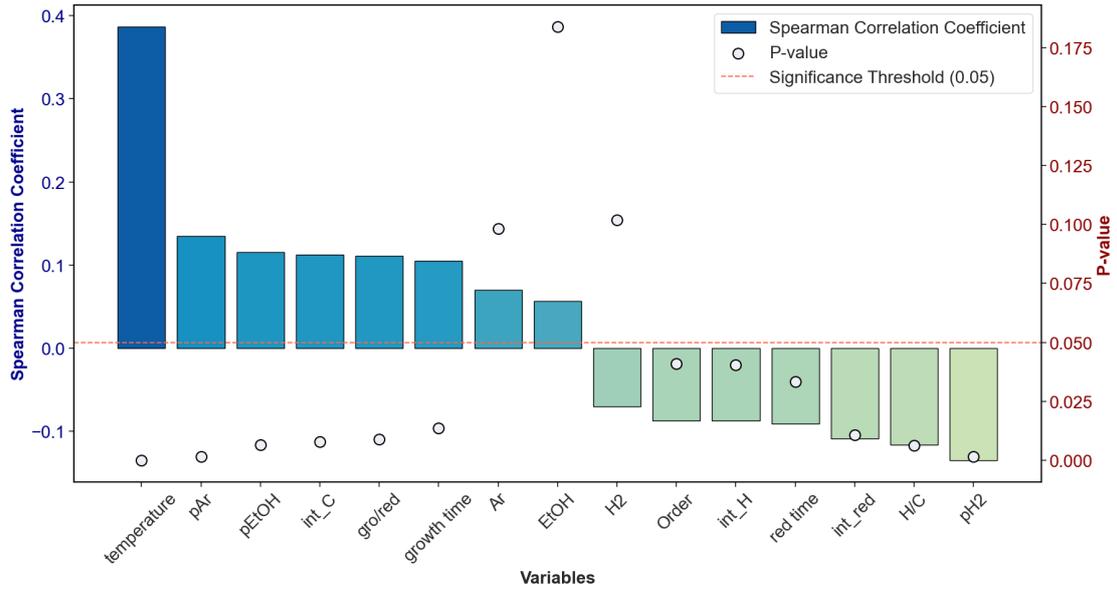

**Supplementary Fig. 5 | Spearman's rank correlation analysis of process parameters.** The effects of temperature, growth time, reduction time, flow rates and partial pressures of Ar, EtOH and $H_2$, H/C ratio (ratio of $H_2$ and EtOH flow rates), int_C (total EtOH flow), int_H (total $H_2$ flow), ratio of growth time to reduction time, and growth order (per day) on the growth outcome (log(Score)) were investigated.

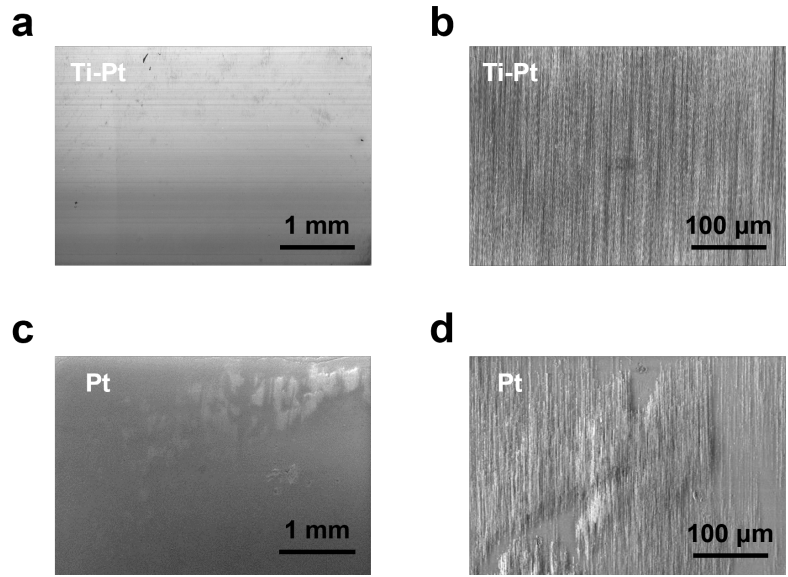

**Supplementary Fig. 6 | Comparison of Uniformity of HACNT Arrays Grown with Ti-Pt and Pt Catalysts.** a-b, SEM images at various magnifications show HACNT arrays grown with Ti-Pt catalysts, demonstrating superior uniformity. c-d, SEM images of HACNT arrays grown with Pt catalysts reveal poor uniformity, however, in specific areas, the growth results resemble those produced with Ti-Pt catalysts.

> *We found that the combination of Ti and Pt, compared to other bimetallic pairs, is more conducive to the growth of high-density HACNT arrays. What do you think the reason is?*

The observation likely stems from several factors related to the catalytic activity, surface energy, and interaction between the metals and carbon during the synthesis process. Here are some considerations that could explain why Ti-Pt bimetallic pairs are particularly effective:

1. **Catalytic Activity**: Both Ti and Pt are known for their catalytic properties, but in combination, they might offer a unique synergy that enhances the decomposition of carbon-containing precursors. This decomposition is critical for the formation of carbon nanotubes (CNTs). Pt is a noble metal with high catalytic activity, especially in hydrogenation reactions, which could facilitate the breakdown of hydrocarbons into carbon atoms for CNT growth. Ti, on the other hand, could help in dispersing the Pt catalyst and stabilizing the active sites.
2. **Surface Energy**: The surface energy of the catalyst particles plays a crucial role in the nucleation and growth of CNTs. The combination of Ti and Pt might result in a bimetallic surface with an optimized surface energy that promotes the formation of CNTs with fewer defects. This could lead to a more uniform growth of high-density HACNT arrays.
3. **Wettability and Interaction with Carbon:** The specific interaction between the Ti-Pt catalyst and carbon might favor the formation of CNTs. Changes in wettability could influence how carbon precursors interact with the catalyst, potentially enhancing the efficiency of CNT growth. The Ti-Pt combination might offer a beneficial environment for carbon diffusion and nucleation, critical for initiating CNT growth.
4. **Diffusion Characteristics:** The diffusion of carbon atoms on and within the catalyst particle is crucial for sustained CNT growth. Ti-Pt bimetallic particles might exhibit optimized diffusion characteristics, allowing for efficient transport of carbon atoms to the growth sites. This can lead to continuous growth of CNTs and the formation of high-density arrays.
5. **Thermal Stability:** The thermal stability of the catalyst is essential for high-temperature CNT growth processes. The combination of Ti and Pt could provide a catalyst with enhanced thermal stability, resisting changes in structure or composition that might deactivate the catalyst.

These factors contribute to a more effective catalytic system for the growth of high-density HACNT arrays, leveraging the unique properties of both Ti and Pt in synergy. Further experimental and theoretical studies would be beneficial to fully understand the mechanisms at play and to optimize the conditions for the growth of high-quality CNTs using Ti-Pt catalysts.

**Supplementary Fig. 7 | Insights from Carbon_GPT on Ti-Pt Catalysts.** This figure delves into the detailed reasoning provided by Carbon_GPT on why the Ti-Pt catalyst combination is particularly effective for synthesizing high-density HACNT arrays. It discusses factors such as catalytic activity, surface energy, wetting and interaction with carbon, diffusion characteristics, and thermal stability, which collectively contribute to the exceptional performance of Ti-Pt in CNT growth.

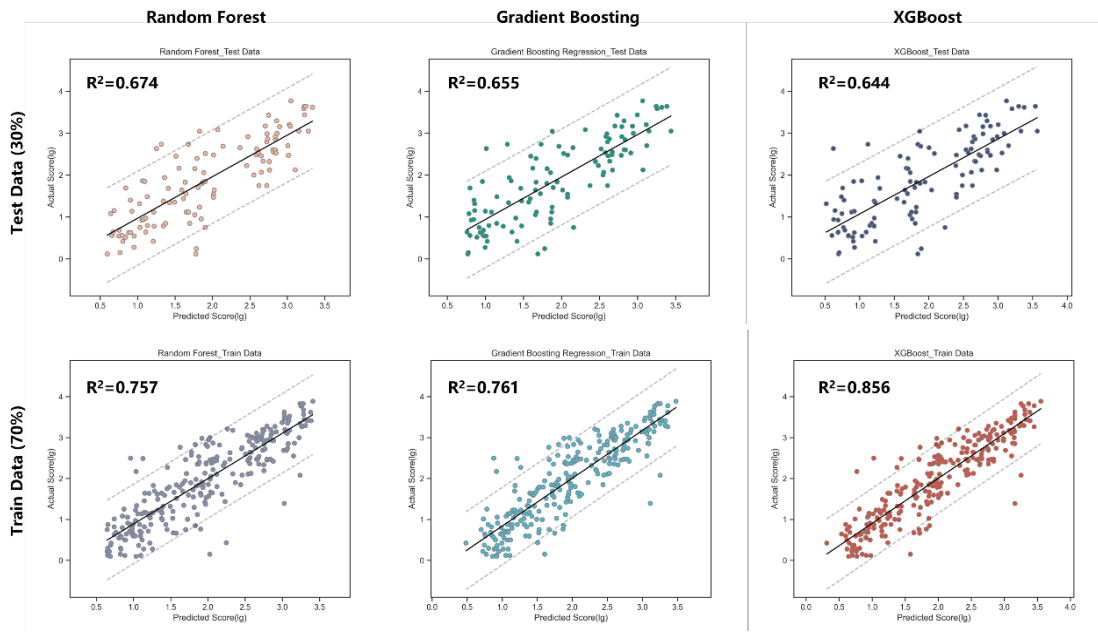

**Supplementary Fig. 8 | Model Performance Comparison.** This figure presents the performance of three machine learning models: Random Forest, Gradient Boosting, and XGBoost. Each model's ability to predict the density scores of HACNT arrays is illustrated through scatter plots, contrasting actual versus predicted scores for both training and testing datasets, with $R^2$ values indicating the level of accuracy for each model.

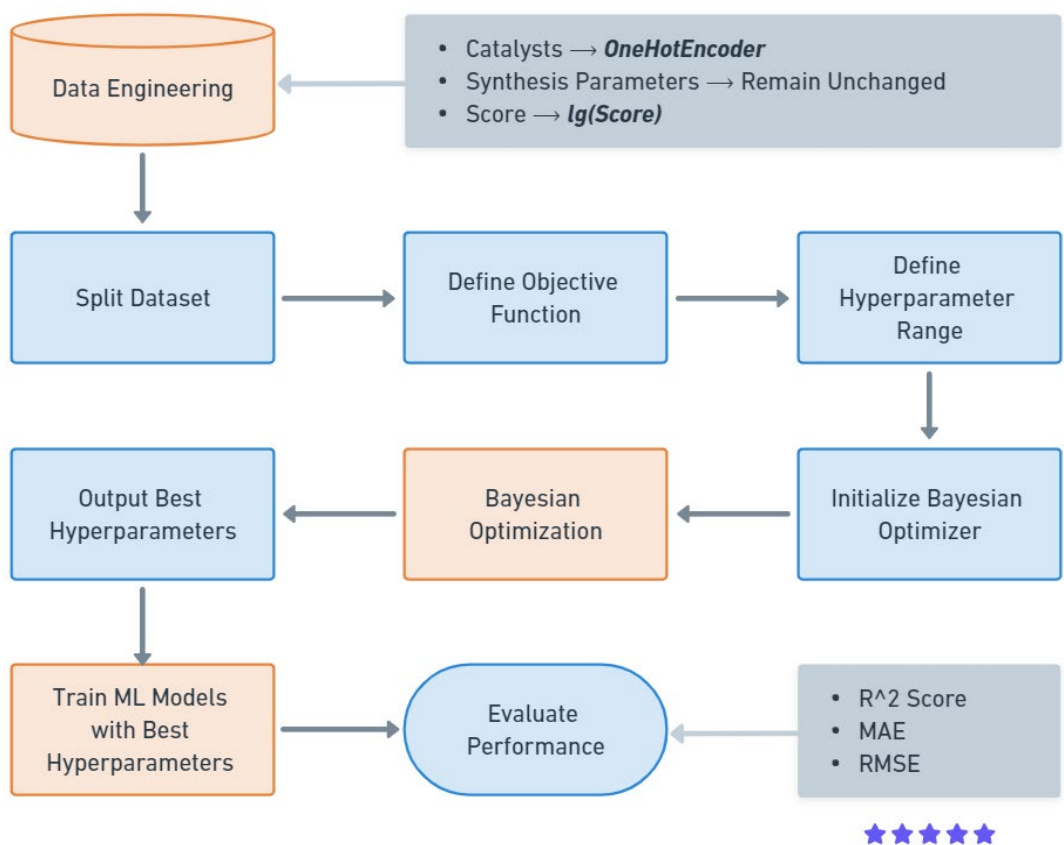

**Supplementary Fig. 9 | Flowchart of Machine Learning Models.** This figure illustrates the process of building machine learning models for HACNT array synthesis. It starts with data engineering, where catalysts are encoded using OneHotEncoder, and the dataset is split. An objective function is defined, followed by the establishment of a range for hyperparameters. Bayesian Optimization is applied to find the best hyperparameters, which are then used to train the machine learning models. The performance of these models is evaluated based on $R^2$ score, Mean Absolute Error (MAE), and Root Mean Square Error (RMSE).

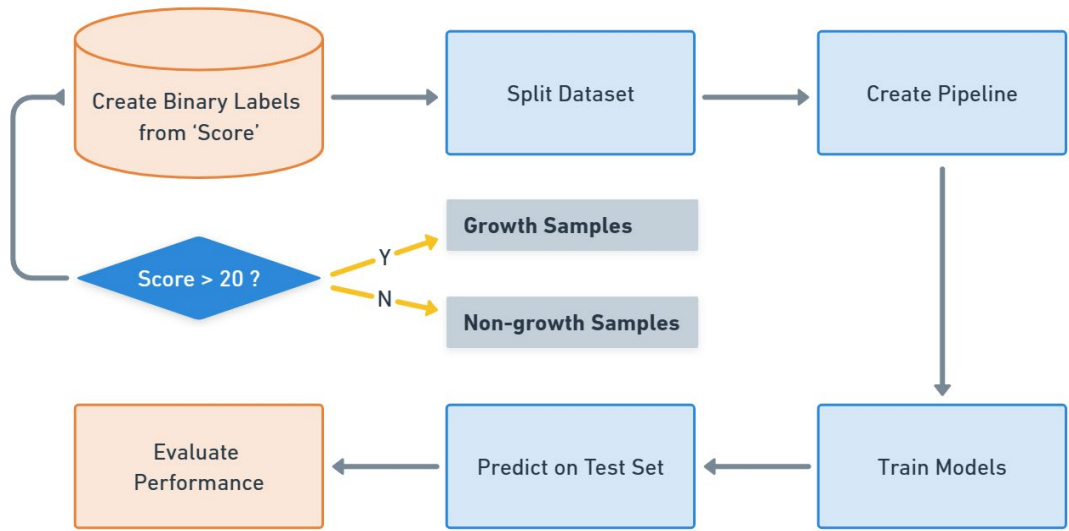

**Supplementary Fig. 10 | Classification Model Training Workflow.** The diagram illustrates the steps taken to train a classification model for distinguishing between growth and non-growth samples of HACNT arrays.

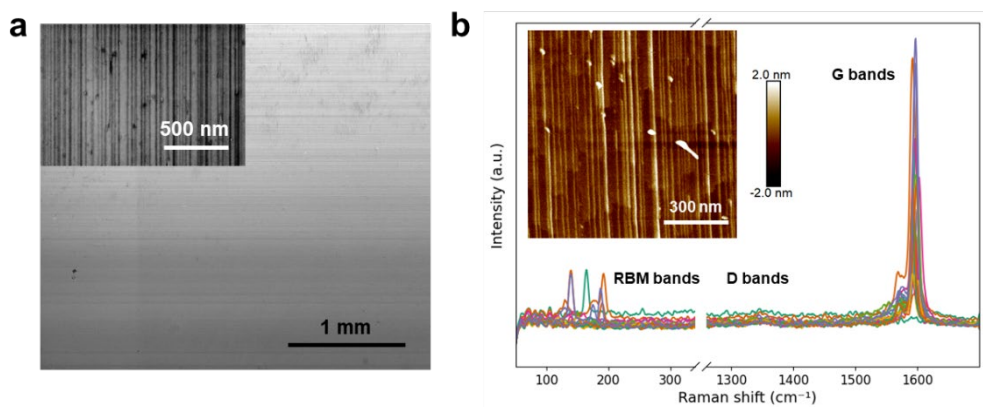

**Supplementary Fig. 11 | General characterization of HACNT arrays.** a, An SEM image captures the macroscopic uniformity of the HACNT arrays grown with Ti-Pt catalysts, with a close-up inset showcasing the microscale density. b, The Raman Spectra (at 532 nm excitation) and AFM image (inset) reveals the good quality and morphology details of the HACNT arrays.

**Supplementary Table 1** | All Recipe Recommendations by CARCO for the Density Controllable Growth Task.

| T/°C | H₂/ sccm | EtOH/ sccm | Ar/ sccm | red_t/s | growth_t/s | lg(score) | expect |
|---|---|---|---|---|---|---|---|
| 907 | 203 | 27.2 | 360 | 339 | 749 | 1.83 | 2 |
| 852 | 217 | 76.7 | 350 | 544 | 772 | \ | 2 |
| 921 | 334 | 62.0 | 342 | 94 | 482 | \ | 2 |
| 931 | 478 | 10.9 | 376 | 380 | 499 | 1.97 | 2 |
| 844 | 336 | 57.7 | 86 | 318 | 1079 | \ | 2 |
| 858 | 207 | 32.7 | 340 | 338 | 741 | \ | 2 |
| 861 | 411 | 49.9 | 82 | 319 | 964 | \ | 2 |
| 921 | 196 | 85.2 | 347 | 343 | 541 | 2.03 | 2 |
| 832 | 210 | 13.7 | 317 | 340 | 728 | \ | 2 |
| 844 | 326 | 88.7 | 320 | 161 | 480 | \ | 2 |
| 933 | 105 | 40.9 | 194 | 179 | 123 | 1.95 | 2 |
| 963 | 416 | 45.9 | 472 | 155 | 280 | 1.72 | 2 |
| 898 | 228 | 34.5 | 485 | 338 | 683 | \ | 2 |
| 894 | 218 | 28.4 | 355 | 338 | 715 | 2.02 | 2 |
| 923 | 222 | 43.3 | 300 | 339 | 646 | 2.49 | 2 |
| 936 | 431 | 105.5 | 82 | 300 | 1149 | 2.10 | 2.5 |
| 936 | 411 | 89.7 | 80 | 310 | 854 | 1.58 | 2.5 |
| 987 | 387 | 60.6 | 108 | 418 | 386 | 2.71 | 2.5 |
| 975 | 458 | 20.1 | 190 | 178 | 319 | 2.45 | 2.5 |
| 973 | 303 | 31.9 | 190 | 172 | 372 | 2.58 | 2.5 |
| 933 | 139 | 8.5 | 430 | 275 | 669 | 2.43 | 2.5 |
| 970 | 495 | 21.5 | 270 | 361 | 541 | 1.96 | 2.5 |
| 914 | 69 | 76.7 | 200 | 64 | 1190 | \ | 2.5 |
| 987 | 387 | 60.6 | 108 | 418 | 386 | 2.17 | 2.5 |
| 845 | 145 | 46.1 | 493 | 214 | 529 | \ | 2.5 |
| 953 | 327 | 29.8 | 429 | 312 | 700 | 2.43 | 2.5 |
| 959 | 67 | 18.9 | 102 | 163 | 1015 | 3.02 | 3 |
| 948 | 411 | 3.5 | 273 | 158 | 905 | 0.88 | 3 |
| 940 | 22 | 47.3 | 427 | 231 | 961 | 1.04 | 3 |
| 952 | 323 | 33.6 | 138 | 144 | 1032 | 1.80 | 3 |
| 948 | 134 | 75.5 | 219 | 521 | 866 | 3.04 | 3 |
| 959 | 119 | 60.1 | 123 | 163 | 985 | 2.80 | 3 |
| 977 | 137 | 6.4 | 427 | 209 | 779 | 2.91 | 3 |
| 954 | 105 | 107.2 | 491 | 223 | 1146 | 1.15 | 3 |
| 950 | 42 | 77.4 | 365 | 292 | 581 | 1.22 | 3 |
| 940 | 101 | 80.0 | 364 | 48 | 682 | 0.28 | 3 |
| 948 | 228 | 17.3 | 173 | 90 | 781 | 2.98 | 3 |
| 946 | 163 | 56.5 | 163 | 319 | 1088 | 3.05 | 3 |
| 940 | 157 | 59.2 | 220 | 501 | 794 | \ | 3 |
| 945 | 137 | 11.1 | 190 | 149 | 849 | 3.52 | 3.3 |
| 945 | 134 | 112.9 | 172 | 136 | 861 | 2.79 | 3.3 |
| 990 | 93 | 82.1 | 292 | 108 | 1164 | 2.26 | 3.3 |
| 952 | 150 | 58.9 | 353 | 165 | 885 | 2.67 | 3.3 |
| 987 | 84 | 102.0 | 373 | 144 | 941 | 2.92 | 3.3 |
| 980 | 81 | 70.5 | 237 | 122 | 1012 | 3.32 | 3.3 |
| 952 | 41 | 68.4 | 292 | 171 | 1065 | 2.82 | 3.3 |
| 982 | 34 | 54.9 | 237 | 133 | 1194 | \ | 3.3 |
| 959 | 95 | 53.2 | 290 | 159 | 961 | 2.83 | 3.3 |
| 953 | 36 | 91.8 | 276 | 152 | 867 | 3.04 | 3.3 |
| 951 | 89 | 4.5 | 299 | 466 | 903 | 3.42 | 3.3 |
| 983 | 105 | 91.3 | 364 | 144 | 1045 | 2.66 | 3.3 |
| 955 | 89 | 23.9 | 370 | 139 | 918 | 3.21 | 3.3 |
| 948 | 111 | 32.7 | 325 | 126 | 794 | 2.87 | 3.5 |
| 947 | 58 | 28.4 | 315 | 122 | 817 | 0.10 | 3.5 |
| 944 | 94 | 10.4 | 311 | 118 | 808 | 3.56 | 3.5 |
| 947 | 65 | 84.2 | 306 | 131 | 1005 | \ | 3.5 |
| 946 | 70 | 13.5 | 283 | 129 | 1173 | 3.39 | 3.5 |
| 945 | 83 | 111.4 | 306 | 132 | 1096 | 1.12 | 3.5 |
| 952 | 23 | 65.8 | 301 | 137 | 846 | \ | 3.5 |
| 945 | 40 | 104.8 | 340 | 141 | 895 | 0.54 | 3.5 |
| 945 | 67 | 99.1 | 332 | 140 | 1188 | 3.32 | 3.5 |
| 952 | 77 | 51.1 | 341 | 142 | 956 | 1.16 | 3.5 |
| 966 | 38 | 25.6 | 320 | 52 | 1057 | 3.02 | 3.5 |
| 966 | 17 | 45.7 | 319 | 32 | 1092 | 0.97 | 3.5 |
| 961 | 101 | 90.4 | 273 | 22 | 1154 | 2.99 | 3.5 |
| 946 | 88 | 10.4 | 289 | 137 | 806 | 3.55 | 3.5 |
| 987 | 18 | 9.0 | 247 | 24 | 859 | 3.31 | 3.5 |
| 984 | 33 | 8.8 | 222 | 39 | 947 | 3.61 | 3.5 |
| 961 | 113 | 8.5 | 322 | 54 | 1134 | 2.61 | 3.7 |
| 954 | 106 | 7.3 | 300 | 132 | 864 | 2.80 | 3.7 |
| 977 | 70 | 3.5 | 314 | 124 | 1120 | 3.77 | 3.7 |
| 982 | 102 | 14.0 | 301 | 129 | 981 | 3.71 | 3.7 |
| 946 | 67 | 6.4 | 329 | 40 | 1056 | 2.67 | 3.7 |
| 955 | 106 | 14.2 | 327 | 126 | 868 | 3.37 | 3.7 |
| 947 | 106 | 4.7 | 321 | 72 | 961 | 3.16 | 3.7 |
| 983 | 110 | 31.7 | 301 | 16 | 910 | 3.18 | 3.7 |
| 951 | 150 | 18.7 | 318 | 13 | 1065 | 3.59 | 3.7 |
| 940 | 142 | 16.6 | 334 | 126 | 1040 | 3.15 | 3.7 |